\documentclass[useAMS,usenatbib]{mn2e}
\usepackage{graphicx}
\title{The influence of spin on jet power in neutron star X-ray binaries}
\author[S. Migliari, J.C.A. Miller-Jones \& D.M. Russell]{S. Migliari$^{1,2}$\thanks{E-mail: smigliari@am.ub.es (SM); james.miller-jones@curtin.edu.au (JCAMJ); D.M.Russell@uva.nl (DMR)}, J.C.A. Miller-Jones$^{3}$ \& D.M. Russell$^{4}$\\
$^{1}$ Departament d'Astronomia i Meteorologia, Universitat de Barcelona, Mart\'i  i Franqu\`es 1, 08028 Barcelona, Spain\\
$^{2}$ Institut de Ci\`encies del Cosmos, Universitat de Barcelona, Mart\'i i Franqu\`es 1, 08028 Barcelona, Spain\\ 
$^{3}$ International Centre for Radio Astronomy Research, Curtin University, GPO Box U1987, Perth, WA 6845, Australia\\
$^{4}$ Astronomical Institute `Anton Pannekoek', University of Amsterdam, P.O. Box 94249, 1090 GE Amsterdam, the Netherlands.\\
}
\begin{document}

\date{}

\pagerange{\pageref{firstpage}--\pageref{lastpage}} \pubyear{2010}

\maketitle

\label{firstpage}

\begin{abstract}

We investigate the role of the compact object in the production of jets from neutron star X-ray binaries. The goal is to quantify the effect of the neutron star spin, if any, in powering the jet.  We compile all the available measures or estimates of the neutron star spin frequency in jet-detected neutron star X-ray binaries. We use as an estimate of the ranking jet power for each source, the normalisation of the power law which fits the X-ray/radio and X-ray/infrared  luminosity correlations L$_{radio/IR}\propto$L$_{X}^{\Gamma}$ (using infrared data for which there is evidence for jet emission). We find a possible relation between spin frequency and jet power (Spearman rank 97\%), when fitting the X-ray/radio luminosity correlation using a power law with slope 1.4; $\Gamma=1.4$ is observed in 4U~1728-34 and is predicted for a radiatively efficient disc and a total jet power proportional to the mass accretion rate. If we use a slope of 0.6, as observed in Aql~X-1, no significant relation is found. An indication for a similar positive correlation is also found for accreting millisecond X-ray pulsars (Spearman rank 92\%), if we  fit the X-ray/infrared luminosity correlation using a power law with slope 1.4. 
While our use of the normalisation of the luminosity correlations as a measure of the ranking jet power is subject to large uncertainties, no better proxy for the jet power is available.  However, we urge caution in over-interpreting the spin-jet power correlations, particularly given the strong dependence of our result on the (highly uncertain) assumed power law index of the luminosity correlations (which span less than 3 orders of magnitude in X-ray luminosity). We discuss the results in the framework of current models for jet formation in black holes and young stellar objects and speculate on possible different jet production mechanisms for neutron stars depending on the accretion mode.

\end{abstract}


\section{Introduction}

Jets are powerful phenomena observed in many astrophysical objects; they are found in both non-relativistic and relativistic systems, including young stellar objects (YSOs), cataclysmic variables (CVs), X-ray binaries (XRBs) and active galactic nuclei (AGN).  In the study of jet production, a major open issue is whether  the mechanism is the same for all these objects. If so, the plethora of information derived from the different types of systems, can, combined, provide important insights into the puzzling problem of jet formation. The accretion disc seems to be essential to the formation of jets in all these systems, with the accretion power being directly related to the outflowing jet power (see below). Cross comparison of the properties of different jet systems is key to understanding the elements contributing to the jet production.

From an observational point of view,  simultaneous X-ray/radio observational campaigns on XRBs already gave us fundamental  information on the physical relations between the accretion disc and the outflowing jet. Correlations have been found between radio and X-ray luminosities in black holes (BHs; e.g. Corbel et al. 2003; Gallo, Fender \&  Pooley 2003; note that recently an increasing number of outliers, the so-called radio-quiet BHs, have been reported: see e.g. discussion in Soleri \& Fender 2011 and Coriat et al. 2011) and in low-luminosity neutron stars (NSs; Migliari et al. 2003; Migliari \& Fender 2006; Tudose et al. 2009; Miller-Jones et al. 2010a) and also between radio and optical/infrared (OIR) luminosities in both BHs and NSs (e.g. Russell et al. 2006 ; Russell \& Fender 2010 for a review on BHs). 

Extending the correlation between radio and X-ray power found for BH XRBs, also including super-massive BHs, and with the addition of the mass parameter, there is evidence for a Ôfundamental plane of BH activityÕ in which a single 3-dimensional power-law function can fit all the BH data (XRBs and AGN) for a given X-ray luminosity,  radio luminosity and mass of the compact object (Merloni et al. 2003; Falcke et al. 2004). These relations link the jet power to a physical quantity in the innermost regions of the accretion disc. 

A unified phenomenological model for the disc-jet coupling in BH XRBs has been proposed (Fender et al. 2004a; Fender et al. 2009).  The scenario is based on the observation that the mass accretion rate variability drives changes in the jet behaviour over the course of an outburst cycle. In particular, the power of the continuous compact jet increases as (or following) the disc mass accretion rate increases, during the rise of the outburst, until the source reaches the hard to soft state transition. At the peak of the outburst, when the source normally reaches a luminosity comparable to the Eddington limit, the compact jet quenches and a transient jet is launched from the system, possibly through internal shock processes (see Fender et al. 2009 for a detailed discussion and caveats).  

A comprehensive study comparing the radio and X-ray  properties of BH and NS XRBs (Migliari \& Fender 2006) has highlighted a number of relevant similarities, such as the fact that below a few per cent of the Eddington luminosity, in their hard state, both BHs and NSs produce steady compact jets (see e.g. Migliari et al. 2006, 2010) while transient jets are associated with highly variable sources or flaring activity at the highest luminosities (e.g. Fomalont et al. 2001; Fender et al. 2004b). Indeed, the hardness-intensity diagram (HID) of transient low-luminosity NS outbursts shows a very similar pattern to that observed in BHs (Maitra \& Bailyn 2004; K\"ording et al. 2008). All these results indicate that the physical mechanism for jet production in accreting NSs and BHs (at all mass scales) is similar. The parameter playing the main role in this mechanism is likely the mass accretion rate. 

However, also significant differences in the jet behaviour of BHs and NSs have been observed. The three main overall differences are the following. There is a difference in radio loudness at the same inferred accretion power (the jets in NSs are less powerful. Moreover, there is a lack of compact jet suppression in at least two NSs when the sources are steadily in their soft state, at luminosities of a few tens per cent of the Eddington limit (Migliari et al. 2004). Note, however, that a possible quenching, similar to that found in BHs, is observed in Aql~X-1 at a few per cent of Eddington (Miller-Jones et al. 2010). Finally, there is a lack of transient optically-thin jets associated with hard to soft state transitions during NS outbursts (Miller-Jones et al. 2010).  These differences may suggest that other parameters related to the nature of the compact object can play a role in powering the jet. 

From a theoretical standpoint, the models for jet production that have been developed during the past years (mainly) focused on two jet systems: BHs, in the case of relativistic systems, and YSOs, in the case of magnetised stars.
  
In BH systems, besides the `extended-disc jet model' of Blandford \& Payne (1982), where the jet is launched by disc magnetic field lines on a wide range of disc radii, another popular jet model is that proposed by Blandford \& Znajek (1977), where the energy is directly tapped from the spin of the BH through disc magnetic field lines anchored to the ergosphere. The fundamental processes of energy extraction described in these models for BHs can in principle be adapted and work also for NSs:  Blandford \& Payne (1982) mentioned it explicitly for the `extended-disc jet model', as the mechanism does not directly involve the accreting star. Furthermore, although the Blandford \& Znajek model cannot be applied as described because of the absence of the ergosphere in NSs, the jet might also directly tap the angular momentum energy of the rotating star by e.g. anchoring the disc magnetic field lines to the magnetised NS itself. 

As far as YSOs are concerned, the three most popular models to explain the observed characteristics of winds/jets\footnote{When referring to YSOs, the terms `disc wind' and `jets from the disc' are often synonymous} are the following: {\it (i)} the {\it extended disc-wind}, where the wind is launched, collimated and accelerated by disc magnetic field lines over a wide range of  disc radii (e.g. K\"onigl \& Pudritz 2000; based on the magnetocentrifugal mechanism proposed by Blandford \& Payne 1982);  {\it (ii)} the {\it X-wind}, whose formation is localized very close to the accreting object, and where the existence of a co-rotational radius is key, and strongly depends on the spin and magnetic field of the accreting star. In the X-wind model, the wind is ejected through the wide-open-angle magnetic field lines of the star which are not threaded into the disc, very close to the corotational radius (e.g. Shu et al. 1994); {\it (iii)} the {\it stellar wind} where the wind is ejected through the open magnetic field lines anchored at the pole of the accreting star.
In the last two of these models, the spin and magnetic field of the accreting object are crucial elements in powering the wind/jet. Adding relativistic corrections, these models can in principle be applied also to NSs. 

In order to quantify the possible relation between the spin and the jet power in BHs, Fender, Gallo \& Russell (2010) recently compared the reported measurements of these two quantities (spin and ranking in jet power) for a sample of stellar-mass and super-massive BH systems. They found no significant correlation. They pointed out the possibility that the reported quantities might not be correct, especially given the uncertainties in the measurements of the BH spin which sometimes differ enormously for the same source depending on the estimating method and the data set used (e.g. in GRS 1915+105 a range of values of the dimensionless spin $a$ between $<0.15$, as in Kato 2004,  and $>0.98$, as in Remillard \& McClintock 2006, have been estimated using different observations).  

Contrary to BHs, the spins of NSs can be directly measured in some cases and estimated with far less uncertainty in others, which allows us to investigate the role of spin in the powering of jets using a better-constrained data set. In this work, we make the first attempt to quantify the role of the spin of the compact object in powering the jets of NS XRBs. We plot the estimated ranking in jet power against the measured/estimated spin of the NS XRBs for which these two quantities are available and then discuss the result in the framework of models for jet production.

\begin{table*}
 \centering
 \begin{minipage}{140mm}
 \begin{tabular}{l l l l l l l l }
  \hline
   Name    & Spin (Hz) & $a$ & Distance (kpc) & $\lambda$ (cm)& Flux density (mJy)   &  &   \\
\hline
&  &  & Atolls  & &   &       &      \\
\hline
4U~1728-34      & $363^{a}\pm5$      &  0.148 &  $5.2\pm0.5$ &  3 & $0.09-0.62$ &       &     \\
4U~0614+091   & $415^{a}\pm5$      &  0.169 & $3.2\pm0.5$ & 6  &$0.08-0.34$  &       &      \\
                             &                          &                                     &    &  3 & $0.13-0.33$        &     & \\
4U1820-30         & $285^{b}\pm65$      & 0.116 &  $7.4\pm0.6$&  6 &   0.13   &       &      \\
                             &                         &   &        &  3 &   0.10   &       &      \\
MXB~1730-335 & $306 ^{a}\pm5$      & 0.125 & $8.8\pm3$ & 6 & $0.29-0.33$        &     & \\
                              &                          &                                     &         & 3 & $0.20-0.21$ &            & \\
Aql~X-1$^{c}$                    &  $550.274^{f}$& 0.224  & $4.6\pm1.4$ & 6  & $0.25-0.46$ &       &     \\
                                 &                   &                              &        & 3  & $0.14-0.68$       &     & \\
\hline
&  &  & Z-type  &  &  &       &      \\
\hline
Sco X-1          & $272^{b}\pm40$ & 0.111   & $2.8\pm0.3$           & 3 & $10^{d}$ &       &      \\
GX 17+2         & $272^{b}\pm50$ &  0.111 & $14\pm2$             & 3 & $1.0^{d}$ &       &      \\
GX 349+2      &  $266^{b}\pm13$&  0.108  & $4\pm0.4$               &  3& $0.6^{d}$ &       &     \\ 
Cyg X-2          & $351^{b}\pm34$ & 0.143  &  $13\pm2$        & 3 &  $0.6^{d}$&       &      \\
GX 5-1            & $288^{b}\pm69$ & 0.117  & $9.2\pm2.7$           &  3 &  $1.3^{d}$&       &     \\
GX 340+0       & $339^{b}\pm8$ & 0.140  & $11\pm3$             &  3&$0.6^{d}$  &       &      \\
XTE J1701-462$^{e}$ & $330^{b}\pm5$  & 0.133  & $8.8\pm1.3$ & 6 & $0.52^{d}$        &     & \\                       
                                     &          &                                     &                  & 3 &  $0.60^{d}$               &     & \\

 \hline
&  &  & AMXPs & &  &       &      \\
\hline
SAX~J1808.4-3658&  $400.975^{f}$& 0.163  & $3.5\pm0.1$  & 13 & $0.3-0.8$ &       &      \\
                                   &                  &  &         & 6 & $0.4-0.8$ &       &      \\
                                   &                  &  &         & 3 & $0.3-0.8$ &       &      \\
IGR~J00291+5934 & $598.892^{f}$ & 0.244 & $5\pm1$      & 6        &     $0.17-0.25$         &     &\\
                                   &                  &  &         &  1.5    &     $1.1$                   &     &\\ 
XTE~J0929-314     & $185.105^{f}$   & 0.075 & $7.8\pm4.2$ &  6      &  $0.31-0.36$         &     &\\ 
XTE~J1814-338    &  $314.357^{f}$   & 0.127 & $6.7\pm2.9$ &  $-$     &  $-$  &       &     \\ 
\hline
\end{tabular}

\caption{Name of the source, spin frequency, dimensionless spin parameter, estimated distance, observational radio wavelength, range of radio  detections. When not indicated otherwise, the spin frequencies are taken from Watts et al. (2008); the distances from Jonker \& Nelemans (2004) and Watts et al. (2008 and references therein); the radio flux densities from Migliari \& Fender (2006 and references therein). } \label{tab_1}
Notes: 
$^{a}$ spin estimated using burst oscillations; 
$^{b}$ spin estimated assuming the spin is equal to the difference between the frequencies of the twin kHz QPOs.
$^{c}$ Coherent millisecond pulsations have been found by Casella et al. (2008), making Aql~X-1 an AMXP. However, all its X-ray properties are consistent with  `normal' atoll behaviour, contrary to the other AMXPs (e.g. Reig et al. 2004; van Straaten et al. 2004);
$^{d}$ Average of the detected radio flux density (Fender \& Hendry 2000; Migliari \& Fender 2006; Fender et al. 2008);
 $^{e}$ 4U~1701-462 showed both atoll and Z-type behaviour, however the radio detections we report here were taken during its Z-type behaviour (Fender et al. 2007); distance from Lin et al. (2007).
$^{f}$ The errors on the spin are small compared to the frequency we show in the Table (of the order of $10^{-4}$ Hz in the case of Aql~X-1 and less than a few $10^{-8}$ Hz for the other AMXPs). Since we do not need such a precision for our purposes, we decided not to show here all the significant digits. 
\end{minipage}
\end{table*}

\section{The sample}

Our sample consists of sources with simultaneous radio and X-ray detections  and with estimated spin and distance. In the following sections, we report on the source sample and the most updated notes on the observations used. The spins, distances and radio flux densities of the sample are shown in Table~1.  We also study a sample of accreting millisecond X-ray pulsars (AMXPs) for which simultaneous X-rays and synchrotron (jet) IR emission are observed. The IR flux densities are shown in Table~2. 

\subsection{The sources: Atolls, Z-type and AMXPs}

Low-magnetic field NS XRBs are commonly divided into two broad classes, based on their X-ray spectral and timing properties, and named after the shape of their color-color diagram (CD): Atolls and Z sources (Hasinger \& van der Klis 1989). 

Atoll sources have luminosities between 0.01 and 0.5 Eddington, while Z sources are persistently above 0.5 Eddington and sometimes are even found at super-Eddington luminosities. In their CD track, atolls show `branches' (X-ray states) where either the soft disc or the hard non-thermal emission dominates the energy spectrum: a hard branch called the island state (the non-thermal emission dominates), and a soft branch called the banana state (the soft disc emission dominates). The Z track typically shows three branches, from softer to harder: Flaring Branch (FB), Normal Branch (NB) and Horizontal Branch (HB).  

4U~1701-462 is the first source showing a transition between Z-type and atoll-type behaviour during an X-ray outburst, going through all the sub-classes and branches within a few months (Homan et al. 2010 and references therein). Homan et al. (2010) showed that the average mass accretion rate (using the 2-2.9~keV count rate as a proxy) seems to be responsible for the different sub-classes,  i.e.  in order of increasing accretion rate\footnote{Note that the classifications have been made following the transitions within sub-classes of a single source. It may happen, when comparing different sources of different subclasses, that e.g. an atoll in the soft state is more luminous than a GX source.} and simplifying their classification to five sub-classes: atolls in their hard state (island), atoll in their soft state (banana), GX sources (a subclass of a few bright NSs, similar to atolls, but persistently at or above $10$ per cent of the Eddington luminosity; see e.g. van der Klis 2006), Sco-like Z sources and  Cyg-like Z sources. Nevertheless, although it seems that variations in the accretion rate can trace the evolution of the CD throughout all the NS sub-classes, the average mass accretion rate alone is not able to account for variations between the different branches {\it within} a sub-class (especially in Z sources), and at least an additional variable seems to be needed (e.g. Homan et al. 2010). 

X-ray analysis of AMXPs showed that their classification in terms of atolls versus Z sources is not straightforward. van Straaten et al. (2007) reported that SAX~J1808.4-3658 and XTE J1814-338  showed systematic differences from atolls (i.e. a shift in the universal scheme of correlations among the frequencies of X-ray quasi-periodic oscillations), whereas sources like XTE~J0929-314 do show the same X-ray properties as those of low-luminosity bursters (atolls at persistently very low luminosities $<0.01~L_{Edd}$; e.g. van der Klis 2006). Interestingly (but even more confusingly), Linares et al. (2007) showed that the X-ray timing properties of IGR J00291+5934 resemble more those of BHs than those of atolls. In this work, we  group and discuss AMXPs as a class of their own. The exception will be Aql~X-1 (see \S~2.2.1).

\subsection{Radio data}

For the radio characteristics of most of the sources in the sample we refer to Migliari \& Fender (2006) and references therein. 

\subsubsection{Atoll sources}

{\it 4U~0614+091: }
The four radio detections of the source are reported in Migliari et al. (2010). The emission is  consistent with being partially self-absorbed synchrotron coming from a slightly variable compact jet.\\

{\it Aql~X-1: }
 The detection of coherent millisecond X-ray pulsations at $\sim550$~Hz was reported by Casella et al. (2008; although visible for only 150 s of the whole RXTE data archive). By definition, the interpretation of this pulsation as the spin of the NS, would make Aql~X-1 an AMXP. However, in our figures we decided to group Aql~X-1 with the atolls rather than AMXPs; this is because the X-ray spectral and timing properties reported in the literature are consistent with `normal' atoll sources such as 4U~1728-34 and 4U~0614+091, and differ from those shown by other AMXPs such as SAX~J1808.4-3658 or  XTE~J0929-314 (as shown in Reig et al. 2004 and van Straaten et al. 2004). Indeed, Aql X-1 has always been referred to as the prototypical atoll source. This choice does not change any of the results of this work and, in fact, it is reflected only in the colour markers of the figures.   
Miller-Jones et al. (2010) report on daily VLA and VLBA observations of Aql~X-1 during the 2009 outburst, the best radio coverage of a NS XRB outburst to date\footnote{JACPOT collaboration: http:\/\/www.astro.virginia.edu/xrb\_jets\/; see also Miller-Jones et al. 2011}. The source was detected 14 times in the radio band, with the radio emission being triggered at the X-ray state transition.  They also observed a radio quenching during the soft state and after the state transition, similar to that found in BHs. The radio emission was always optically-thick along the whole X-ray outburst track, indicating that it arose from a compact jet.  In this work we use all the reported radio/X-ray observations of the source as collected in Tudose et al. (2009) and Miller-Jones et al. (2010). \\

{\it 4U~1728-34 and 4U~1820-34:}
The radio detections of 4U~1728-34 are taken from Migliari \& Fender (2006). When dual band radio observations were available, the radio spectra were always consistent with being flat, consistent with emission from a compact jet.  4U~1820-34 has been detected in the radio band when it was steady in its soft state (banana; Migliari et al. 2004). The radio luminosity is comparable to that observed in the hard state from other sources. This is different from what has been observed so far in BHs, whose soft state radio emission is at least an order of magnitude lower than that in the hard state. The radio spectrum was consistent with being optically thick, indicating a compact jet. \\

{\it MXB~1730-335 (the Rapid Burster):}
The source has been detected during two X-ray outbursts (Rutledge et al. 1998; Moore et al. 2000).  Contrary to what is expected for BHs, but consistent with what is observed in Aql~X-1, the radio spectra of the two dual-band observations have slopes of $+0.7\pm0.3$ for the 1996 outburst, and of $0.85\pm0.6$ for the 1997 outburst, consistent with a compact jet.

\subsubsection{AMXPs}

Only three out of the 12 known AMXPs (not including Aql~X-1) have been detected in the radio band, all during X-ray outbursts (SAX~J1808.4-3658, IGR~J00291+5934 and XTE~J0929-314). 
IGR~J00291+5934 was detected three times during the 2004 outburst (Pooley 2004; Fender et al. 2004c). For SAX~J1808.4-3658 we plot radio detections of three different outbursts, the one in 1998 (Gaensler, Stappers \& Getts 1999), the one in 2002 (Rupen et al. 2002a) and the one in 2005 (Rupen et al. 2005);  the simultaneous multi-band radio observations available show a flat spectrum, indicating emission from a compact jet.  XTE~J0929-314 was detected twice during the 2002 outburst (Rupen et al. 2002b), but  the radio spectrum is not constrained.

\subsubsection{Z-type neutron stars}

Eight\footnote{Nine if we also count GX13+1, which is a hybrid atoll-Z source} highly-accreting Z-type XRBs are known: six persistent (Sco X-1, GX 17+2, GX 349+2, Cyg X-2, GX 5-1, GX 340+0),  one transiting from atoll to Z-type (XTE J1701-462) and the peculiar source Cir X-1: all of them have been detected in the radio band. The radio emission in Z souces is extremely variable and seems to be related to the X-ray state. Strictly simultaneous radio/X-ray studies  have been reported for six Z sources: GX~17+2 (Penninx et al. 1988; Migliari et al. 2007), Cyg~X-2 (Hjellming et al. 1990a), Sco X-1 (Hjellming et al. 1990b; Fomalont et al. 2001; Bradshaw et al. 2003), GX 5-1 (Tan et al. 1992), XTE~J1701-462 (Fender et al. 2007) and Cir X-1 (Soleri et al. 2009). As in Migliari \& Fender (2006), we are interested in the X-ray/radio luminosity normalisation factor, therefore we use averaged luminosities. To their sample of Z sources we add the averaged radio flux density of XTE~J1701-462 (Fender et al. 2007) using a distance of 8.8~kpc (Lin et al. 2009). 

\begin{table}
 \centering
  \caption{$K$-band flux densities, either detected or extrapolated, of jets from AMXPs. The table shows the name of the source, the date of the detection, the $K$-band flux densities in mJy, the actual filter used to infer the $K$-band jet flux. When the $K$-band  flux density was extrapolated the errors on the flux densities reflect the uncertainty of the slope of the synchrotron jet spectrum (see \S~\ref{sec_oir}). }
 \begin{tabular}{l l l l}
  \hline
   Name   & Date (MJD) & $K$-band (mJy)& Obs. \\
   \hline
SAX~J1808.4-3658 & 50921&   $2.077\pm0.057$& $K$ \\
                                     & 53527 & $0.47\pm 0.30 $ & $I$ \\
IGR~J00291+5934  & 53348& $0.389\pm 0.052 $  & $K$ \\
                                     & 53349 & $0.302\pm 0.010 $ & $K$\\
                                     & 53350 & $0.295\pm 0.014 $ & $K$\\
                                     & 53351 & $0.263\pm 0.023 $ & $K$\\
                                     
                                     & 54694& $0.465\pm 0.250 $  &  $y$\\
                                     & 54729& $0.370\pm 0.052 $&$K$ \\
                                     & 54732& $0.546\pm 0.063 $  &$K$\\
                                     & 54735& $0.348\pm 0.053 $ & $K$\\
                                     & 54740& $0.217\pm 0.038 $  & $K$\\
                                 
XTE~J0929-314       & 52398 & $0.067\pm 0.043 $  & $I$\\
                                     & 52405 & $0.18\pm 0.11 $  & $I$\\
                                     & 52413 & $0.108\pm 0.069 $  & $I$\\
                                     & 52431 & $0.034\pm 0.022 $   & $I$\\
                                   
XTE~J1814-338       & 52814 & $0.31\pm 0.20 $& $I$\\

\hline
\end{tabular}

\end{table}

\subsection{Optical-IR data}
\label{sec_oir}

We take observations of systems with evidence for a NIR-excess above what is most likely the accretion disc component in the optical.  These have been attributed to synchrotron emission from the jet. An optical/IR spectral energy distribution (SED) is required (fluxes acquired in several filters) in which the power law slope of either the disc or jet components has been measured. A quasi-simultaneous X-ray flux (2--10 keV) is also required. For a compilation of optical/IR SEDs where this NIR-excess is evident, see Russell et al. (2007), and for additional recent observations see Torres et al. (2008) and Lewis et al. (2010).

We use the jet flux density measured in the same waveband for each source; the NIR $K$-band (2$\mu$m) in which the jet is typically much brighter than the disc component.  Some observations contained $K$-band fluxes, which we take directly as the jet flux density (IGR J00291+5934 from Torres et al. 2008 and Lewis et al. 2010 and SAX J1808.4--3658 from Wang et al. 2001). For sources in which no $K$-band observations were made, we infer the $K$-band flux density of the jet from the SED. All the flux densities are de-reddened to account for interstellar dust extinction. For some observations of XTE J0929--314 (Giles et al. 2005), SAX J1808.4--3658 (Greenhill et al. 2006), IGR J00291+5934 (Lewis et al. 2010) and XTE J1814--338 (Krauss et al. 2005), optical SEDs were obtained only, for which we take the flux density from the reddest (lowest frequency) filter; i.e. the filter with the most prominent excess ($I$-band or $y$-band at 1 $\mu$m).  In these cases, we subtract the best power law fit to the disc component as inferred at higher frequencies (typically from R, V and U-band observations), in order to estimate the flux from the jet component only.   The jet spectrum in optical/IR could be optically thick, approximately flat ($\alpha \approx 0$) or slightly inverted ($\alpha \approx +0.3$), or optically thin ($\alpha \approx -0.7$). To take into account the uncertainties in the jet spectral index we assume $-1.0 < \alpha < +0.5$ to infer the $K$-band flux density of the jet and the error. We show the observed or derived $K$-band flux densities in Table~2.

\subsection{Measuring the spin of the neutron star}
\label{sec_spin}
Unlike in BHs, the spin in some NS XRBs can be directly observed via non-thermal emitting particles accelerated by the polar magnetic field of the rotating NS. This is the case for high-magnetic field  XRB pulsars ($B>10^{11}$~G) and AMXPs, where a coherent pulsation is directly detected in the X-ray power density spectrum. 

In certain low-magnetic field NS XRBs, the spin periods can be observed indirectly as  `burst oscillations', i.e. X-ray quasi-coherent oscillations during type-I X-ray bursts. In the three sources where both direct pulsations and burst-oscillations have been observed, the burst oscillation was consistent with the spin frequency (Chakrabarty et al. 2003; Strohmayer et al. 2003; Altamirano et al. 2008). As a caveat, note that the burst oscillation period has been observed to drift slightly between observations and the difference between the two frequencies can in fact be up to several Hz. 

Twin kHz quasi periodic oscillations (QPOs) in the X-ray power spectra have been observed for a number of NS XRBs. The difference between the peak frequencies $\nu_{kHz}$ of the two QPOs seems to be in some relation with the spin frequency ($\nu_{s}$; either consistent with or half $\nu_{s}$ depending on the source; see van der Klis 2006 for a review). In some cases, a decrease in $\Delta\nu_{kHz}$ at high and low accretion rates has been observed, which further complicates the direct association between these two quantities (e.g. van der Klis 2006). 
In this work, for those sources where a direct measurement of the pulsation of the NS is not available, we choose to set the spin of the NS equal to the average burst oscillation frequency or the highest $\Delta\nu_{kHz}$ measured. 
 
In order to directly compare our NS sample with that of BHs (from Fender et al. 2010), we estimated also for NSs the dimensionless spin parameter $a=c/GM^{2}\times I\omega$ (where $I$ is the moment of inertia and $\omega/2\pi=P$ is the spin frequency), assuming for all the NSs in the sample a mass M$_{NS}=1.4$~M$_{\odot}$ and a radius R$_{NS}=10$~km. In Table~\ref{tab_1} we show the spin frequency and the relative $a$ values for the sample. 

For the spin measurements and distances we refer to the compiled tables in Watts et al. (2008 and references therein) and Jonker \& Nelemans (2004). The spin of XTE~J1701-462 is estimated to be equal to the difference in frequency of the twin kHz QPOs, observed by Sanna et al. (2010). The separation of the kHz QPOs in GX~340+0 is from Jonker et al. (2000).

\section{Estimating the jet power ranking}
\label{sec_jet_pow}
Following Fender, Gallo \& Russell (2010), we estimate the {\it ranking} jet power of the sample, using as a proxy the normalisation of the power law fitting the sources in the X-ray/radio and X-ray/infrared luminosity planes. 
This ranking method relies on two main assumptions. We assume (1) that the relation between soft X-ray luminosity L$_{X}$ (proxy of the mass accretion rate) and the radio/IR flux density L$_{R/IR}$ of the compact jet (proxy of the jet power) is the same for all the low-magnetic field NS XRBs.
This assumption implies that if the L$_{R/IR}$-L$_{X}$  correlation is a power law and no other parameter plays a role in powering the jet, all the NS sources would lie on the same power-law. Therefore, assuming (2) that an extra parameter affects the powering of the jet, this parameter would shift the L$_{R/IR}$-L$_{X}$ relation above or below that of a pure accretion-jet relation for a jet with a small Lorentz factor ($\Gamma<2$). Note that the compact jets in NSs are observed to be in general weaker than the jets in BHs  and no beaming effect is supposed to play a role in the normalisation of the disc-jet relation (e.g Fender \& Hendry 2000). 

As already discussed in Fender et al. (2010) there are significant uncertainties associated to this ranking method for BHs. In fact, these uncertainties are even larger in the case of NSs. This is for two main reasons. First, the X-ray/radio luminosity correlations found for the atolls during their hard state are constrained over less than three orders of magnitude in X-ray luminosity (compared to  the $6-7$ orders of magnitude for stellar BHs). Second, two different correlations with two different slopes have been found for different NS sources (see \S~\ref{sec_x-r_lum}). Therefore, this represents the major source of uncertainty in our results. Future EVLA deeper radio observations will be able to extend these correlations to lower luminosities and to constrain them with much more accuracy.  Nevertheless, for the present, this is the best proxy for the jet power available.

The existence of the three-dimensional fundamental plane of BH activity (i.e. mass, X-ray luminosity and radio luminosity; Merloni et al. 2003; Falcke et al. 2004) indicates that the mass of the compact object also plays a role in the dispersion of the X-ray/radio luminosity correlations in BHs (as $L_{R}\propto M^{0.8}$). However, in the case of NSs where the measured dispersion in the mass distribution is small, this effect should not be significant [e.g. Valentim, Rangel, Horvath (2011) observed a bimodal mass distribution with a main, very narrow $\sigma=0.042$ peak at $\sim1.4~M_{\odot}$, and a second $\sigma=0.25$ peak at $1.7~M_{\odot}$].  

\begin{figure*}
\begin{tabular}{c}
\includegraphics[scale=0.45]{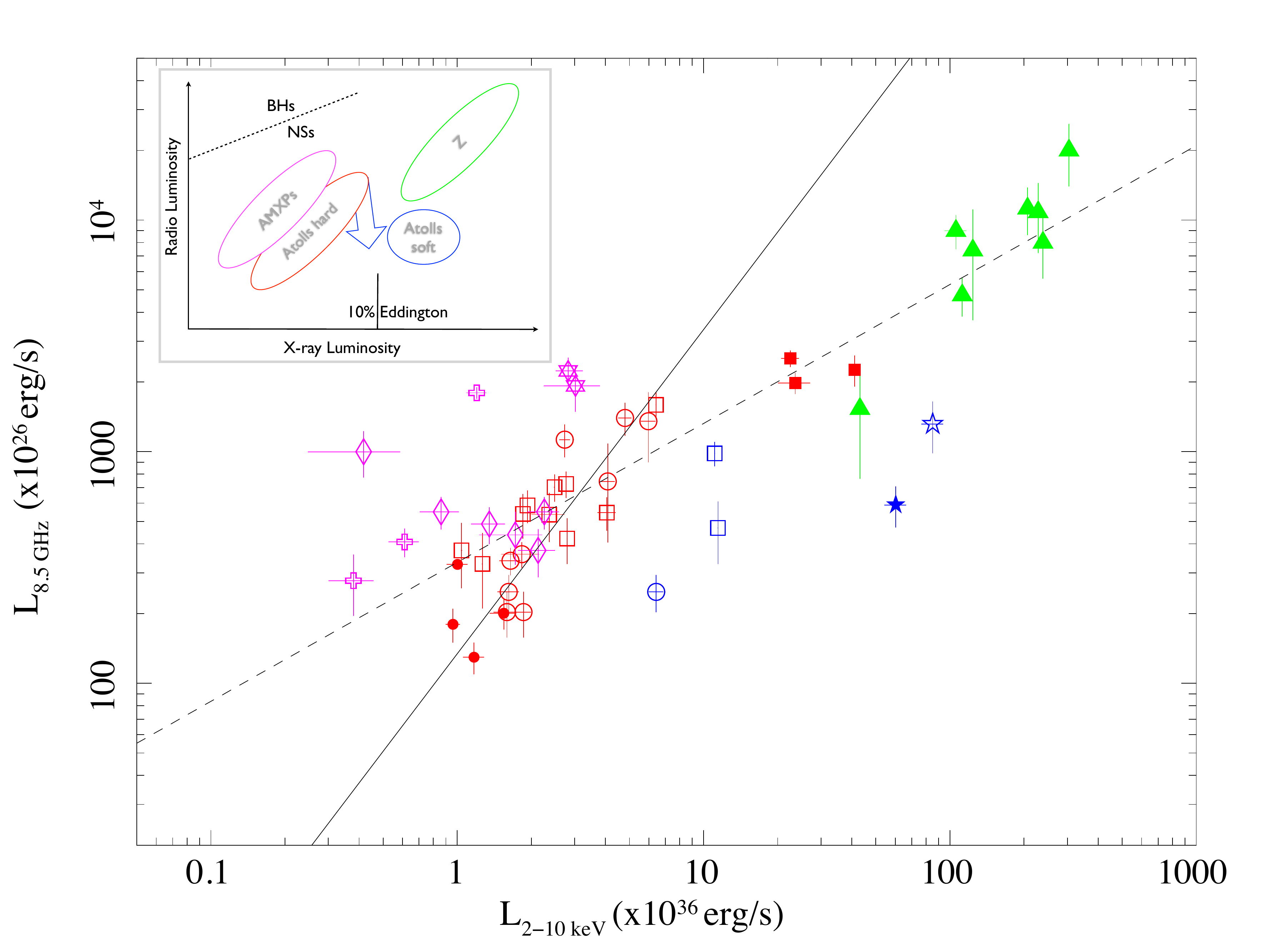}\\
\end{tabular}
\caption{X-ray vs radio luminosity of NS XRBs. The solid line represents the power-law fit of 4U~1728-34 with $\Gamma=1.4$. The dashed line represents the power-law fit of Aql~X-1 with $\Gamma=0.6$. The inset at the top-left corner shows a visualisation of the different classes of NSs in the sample. Markers: open circles, 4U~1728-34; filled squares, MXB 1730-335; open squares, Aql X-1; filled circles, 4U~0614+091; open star, Ser X-1; filled star, 4U~1820-30;  filled triangles are Z sources; open diamonds,  SAX~J1808.4-3658; open crosses are IGR~J00291+5934; stars of David, XTE~J0929-314. Colors: red, atolls in hard state or in outburst, up to the peak;  blue, atolls in steady soft state; green, Z sources; magenta, AMXPs.}\label{fig_r-x_corr}
\end{figure*}

In Fig.~\ref{fig_r-x_corr}, we present the most up to date radio/X-ray luminosity plot with simultaneous X-ray and radio observations of NSs. Note that we plot only radio detections. In Fig.~\ref{fig_ir-x_corr}, we show the IR/X-ray luminosity plot of four AMXPs, as discussed in \S~\ref{sec_oir}.\\ 

%############# FIGURE ##########################

\begin{figure*}
\begin{tabular}{c c}
\includegraphics[scale=0.3,angle=90]{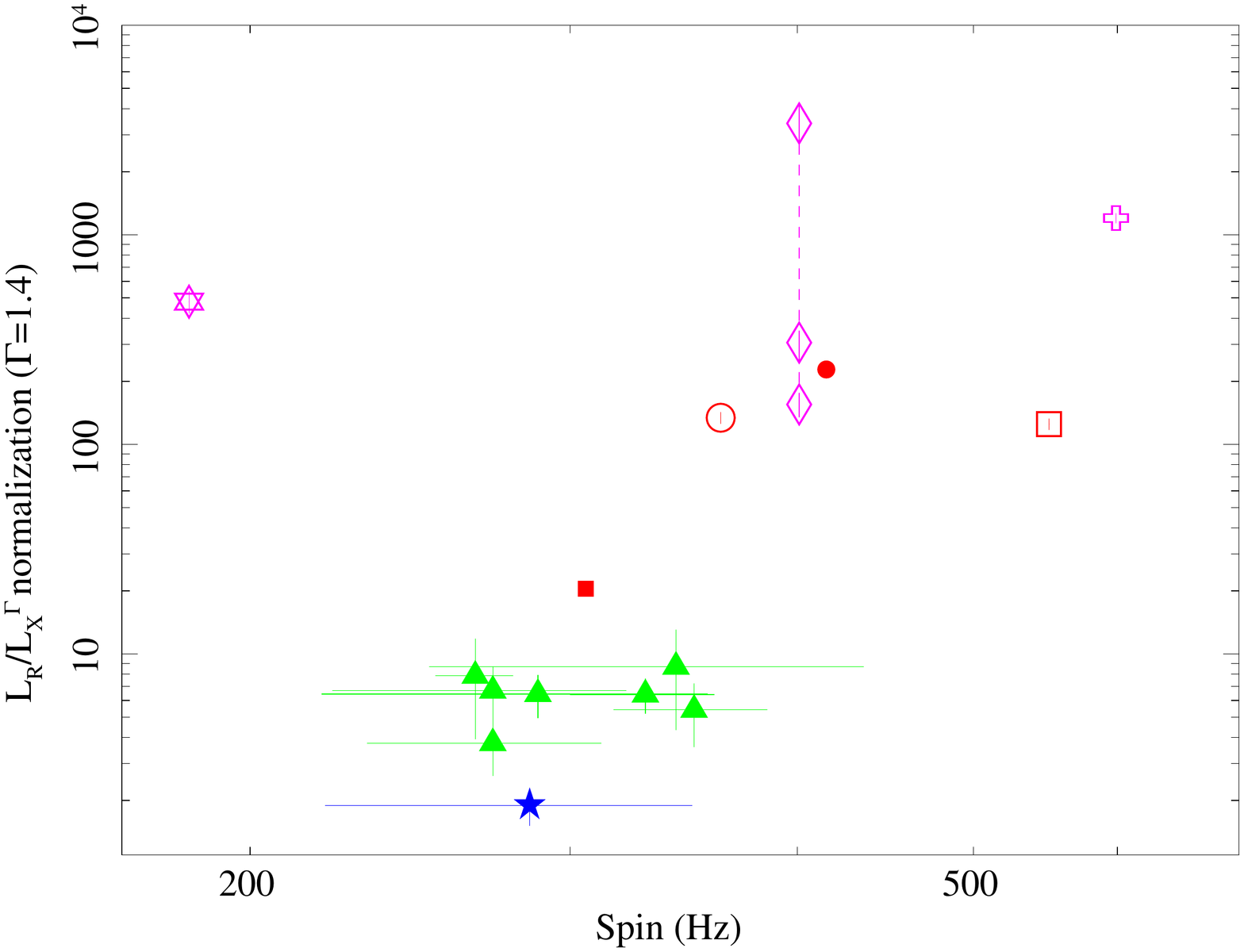}&
\includegraphics[scale=0.3, angle=90]{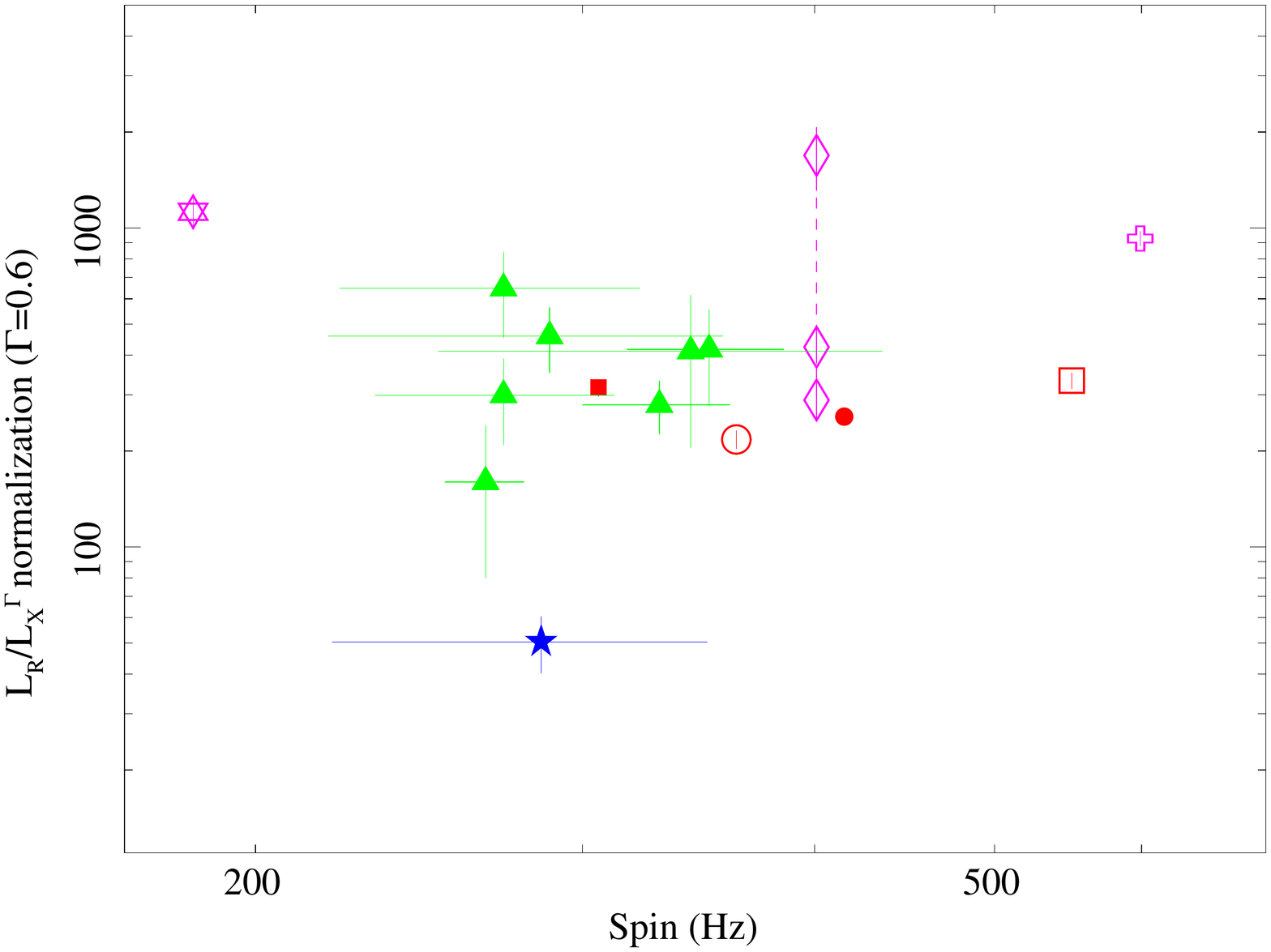}\\
\end{tabular}
\caption{Spin of the NS versus the normalisation of the power law fitting the radio/X-ray luminosity plot, as a proxy of the relative jet power, for each of the sources of the sample. Left panel: the slope of the power law was fixed to $\Gamma=1.4$. Right Panel: the slope of the power law was fixed to $\Gamma=0.6$. Markers as in Fig.~\ref{fig_r-x_corr}. The vertical dashed line joins estimates of the normalisation for different outbursts of the same source (SAX~J1808.4-3658). }
\end{figure*}

%###############################################

\subsection{X-ray/radio luminosity}
\label{sec_x-r_lum}

All the radio spectra observed in atoll sources are consistent with being from optically-thick, partially self-absorbed synchrotron radiation, i.e. a compact jet. There is no confirmed observation of a transient relativistic blob-like ejection like those observed in BH systems. Radio lobes have been resolved only in Sco~X-1 and Cir X-1, both still consistent with the interaction of the jet and the interstellar medium (Fomalont et al. 2001; Fender et al. 2004; Tudose et al. 2009). 
Therefore, we choose to treat the NS radio emission as, on average, coming from a compact jet; the compact jet would dominate the spectra of AMXPs, atolls and the average spectrum of Z sources.

As far as atoll sources in the hard state are concerned, only two, 4U~1728-34 and Aql X-1, show a correlation in the radio/X-ray luminosity plane, with a slopes of $\Gamma\sim1.4$ and $\Gamma\sim0.6$ respectively (Migliari et al. 2003; Tudose et al. 2009; Miller-Jones et al. 2010a). This is due to a lack of radio observational coverage of other atoll sources and their instrinsic faintness in the radio band and hence a smaller lever arm to accurately determine the correlation index. A power law with $\Gamma\sim0.6$ is consistent with what is observed in BHs. The dataset of Aql~X-1 was taken from radio/X-ray detections of the source during an X-ray outburst, whose HID track resembles the q-shape observed in BHs (Miller-Jones et al. 2010a). On the other hand, the observations of 4U~1728-34 ($\Gamma\sim1.4$) were taken during one of the frequent upper-island to the lower-banana state transitions, therefore it does not show the entire q-track as in Aql~X-1 and it may look more like a bi-modal correlation (Migliari et al. 2003).  

Theoretically, given a total jet power proportional to the mass accretion rate, we expect a slope of 1.4 for radiatively efficient systems such as NSs and 0.6 for radiatively inefficient systems like BHs in the hard state (see K\"ording et al. 2006, 2007; see also Coriat et al. 2011 for a 1.4 slope in a BH system). 

More data from atolls in their hard states are necessary to quantify the correct coupling between radio and X-ray luminosity, if, indeed, a power law with the same slope can describe the correlation of all the sources.  In this work, we fit each source in the X-ray/radio luminosity plane with a power law fixing the slope first to 1.4 and then to 0.6, and report the normalisation for both. 

\subsection{X-ray/IR luminosity}

Like BHs (e.g. Corbel \& Fender 2002, Coriat et al. 2009), an optical/IR excess above the accretion disc flux has been reported for some NS XRBs, and identified as likely synchrotron emission from the jet (Wang et al. 2001; Krauss et al. 2005; Giles et al. 2005; Greenhill et al. 2006; Migliari et al. 2006; 2010; Russell et al. 2007; Torres et al. 2008; Lewis et al. 2010). These are generally systems with short orbital periods and hence smaller and dimmer accretion discs, comprising the AMXPs and the ultracompacts.  We can therefore use the optical/IR luminosity of the jet to also infer the jet power in the same way as we use the radio flux of the jet.

\section{Results}

%################## FIGURE ###########################

\begin{figure}
\includegraphics[scale=0.7]{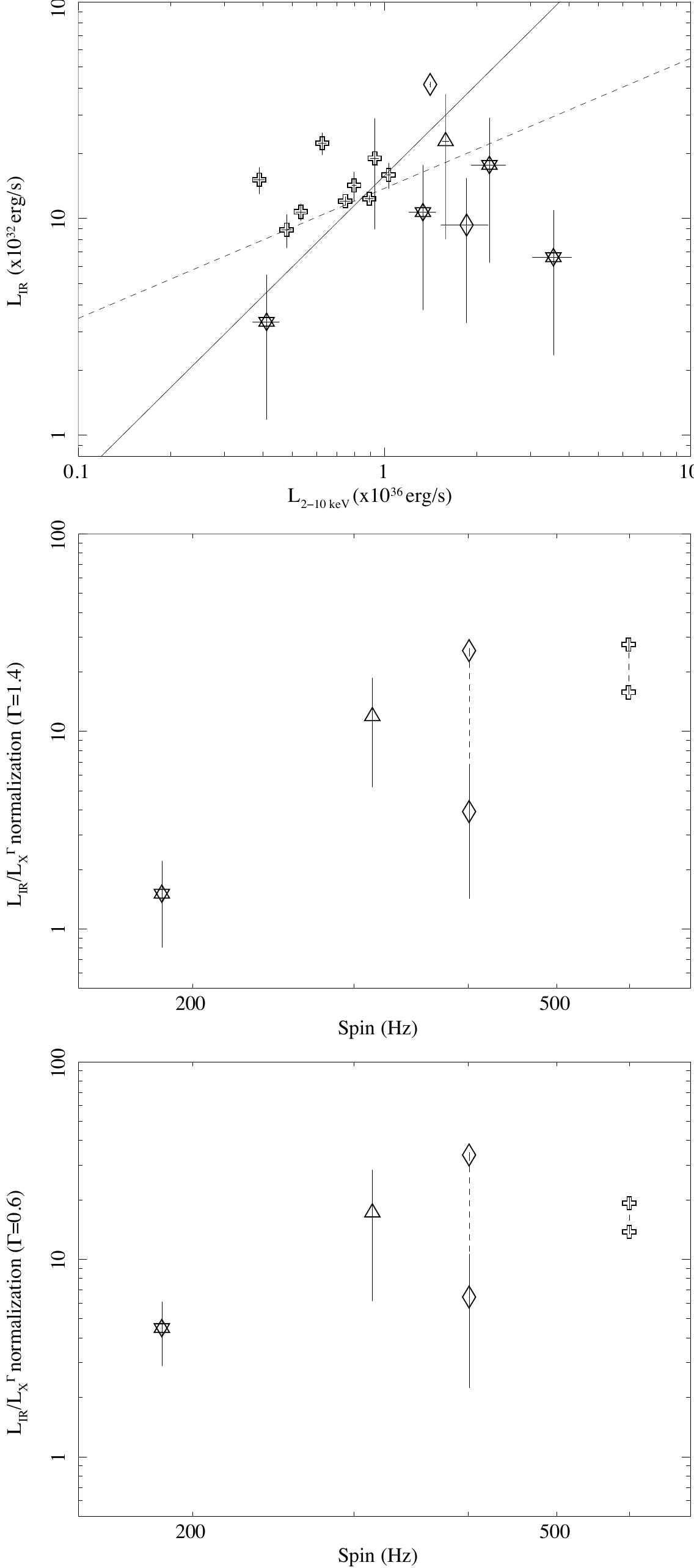}
\caption{In the top panel, we show the X-ray versus IR luminosity in AMXPs. The IR emission represents only the jet excess contribution, when the disc emission is subtracted (see \S~\ref{sec_oir}). The dashed and solid lines represent the power law fit with slope fixed to 0.6 and 1.4, respectively, and normalised to the 2004 outburst of IGR~J00291+5934. The middle and lower panels show the spin estimates plotted against the normalisation of the power law fitting the IR/X-ray luminosity of the sources, fixing the power-law slope to $\Gamma=1.4$ and $\Gamma=0.6$, respectively. Markers as in Fig.~1; here, the open triangle represents XTE~J1814-338. The vertical dashed lines join estimates of the normalisation for different outbursts of the same source (SAX~J1808.4-3658 and IGR~J00291+5934).}\label{fig_ir-x_corr} 
\end{figure}

%#######################################################

The rank in jet power L$_{Jet}$ among the NSs, estimated as the rank in the normalisation of the relation  L$_{R}\propto$L$_{X}^{\Gamma}$ with $\Gamma=1.4$ (i.e. the fit to 4U~1728-34 and as expected theoretically for radiatively efficient accretion when $L_{Jet}\propto \dot{M}$),  shows an apparent correlation with the spin of the sources (Fig.~2): the non-parametric Spearman rank statistic gives a probability of $3\times10^{-2}$ against the null hypothesis of an uncorrelated data set. However, given the outlier (XTE~J0929-314) and the uncertainties discussed above (\S~\ref{sec_spin} and \S~\ref{sec_jet_pow}), much better statistics is necessary to confirm the relation.  
If we plot the spin frequency against the normalisation fixing the power law relation to L$_{R}\propto$L$_{X}^{0.6}$ (fit to Aql~X-1), there is no obvious correlation. 

In Fig.~\ref{fig_ir-x_corr}, we observe a barely significant positive correlation between spin and jet power as inferred from IR jet flux for AMXPs in the case of $\Gamma=1.4$ (Spearman rank 92\%, taking the average normalisations for SAX~1808.4-3658 and IGR~J00291+5934), whose significance decreases with the choice of  $\Gamma=0.6$. 

We note that the indication of a ranking correlation between spin and jet power despite the uncertainties is at least intriguing and definitely worth further studies, especially because of the important implications that such correlations  would have on our understanding of the jet formation mechanisms.

\section{Discussion}

In order to quantify the role of the spin in powering the jet in XRBs, we investigated the relation between the spin and the (ranking) jet power of a number of NS XRBs. We found that, in the case of an X-ray/radio luminosity power-law correlation with slope 1.4, there is a slightly significant (97\%) ranking correlation between spin and the power-law normalisation. No significant correlation is observed if we fit the X-ray/radio luminosity relation using a power law with slope 0.6.  
As a caveat, we remind the reader that the distance, as well as the estimate of the spin of Z sources based on the kHz QPO frequency difference adds uncertainties to the correlation.
A similar correlation is observed in AMXPs, however, between spin frequency and the jet power ranking estimated using IR observations, where the spin is precisely measured. 

Comparing the results from studies of NSs and BHs, we envisage the following possible scenarios.

\begin{enumerate}
\item We know that the jet power of NSs as a function of the mass accretion rate is of the same order of magnitude as BHs (modulo a factor a of a few;  K\"ording, Fender \& Migliari 2006). The spin values in our sample of NSs are $a<0.25$. 
Let us assume that the jets from BHs are formed by efficiently tapping energy from the spin of the compact object. Blandford \& Znajek (1977) predict a quadratic spin dependence on jet power for BHs (L$_{Jet}\propto a^{2}$). 
If also NSs can efficiently extract energy from the spin of the compact object, albeit with a mechanism which would not involve general relativity and in particular the presence of an ergosphere, e.g. through self-collimated stellar winds or anchoring the disc to the NS directly through the stellar magnetic field {\it a la} X-wind (Shu et al. 1994) or Reconnection X-wind (Ferreira et al. 2000; see Ferreira et al. 2006a for a discussion of different possible magnetospheric-wind configurations in YSOs), then there are two possibilities: either the dependence of the jet power on the spin is similar in NSs and BHs and the BHs are all slowly rotating, spinning at a rate comparable with that of our NSs sample, or the spin of the BHs are higher than that of our NSs sample, but the mechanism of extraction of energy from the spin in NSs is much more efficient than that in BHs. 
\item The mechanism for jet production is the same for both NSs and BHs, but it is related to the disc only (e.g. Blandford \& Payne 1982), the compact object having only a mild effect on the jet power. In this case, the indications for the spin-jet power correlations observed in NSs would be an effect of the poor statistics or of incorrect assumptions in the estimate of the jet power ranking. 
\item The mechanism for jet production is completely different for NS and BH systems. The mass accretion rate is correlated to the jet power in both systems, but the actual mechanism of extracting energy and launching of jets is dependent on the compact object. In this case, the relation found between spin and jet power might indicate that the spin and the magnetic field have an active role in powering the jet in NSs, as predicted by e.g.  propeller effect, X-wind, stellar wind models. Note that, because of our lack of information about the magnetic field strength of most of the NSs in our sample, the interplay between the spin and the magnetic field and their specific roles in the jet production cannot be properly disentangled. This would cause a blurring in any existing intrinsic correlation between spin and jet power. 
\item The mechanism for jet production depends on the accretion regime, i.e. on the X-ray state and mass accretion rate. This possibility would introduce an additional variable in the attempt to quantify the role of the different parameters of the system in producing jets.  If this is the case, the mass accretion rate is related to the jet power, but, depending on the accretion mode, the inner disc interacts more or less with the boundary layer of the compact object and a different `jet regime' may start. 
\end{enumerate}

In the framework of the jet models developed for YSOs, the lack of evidence of a strong relation between spin and jet power in XRBs would suggest that the main driver of winds/jets is not strictly linked to the accreting object. This would point towards the extended-disc model as the most likely mechanism. The same conclusion for YSOs  has been derived by Anderson et al. (2003) and Ferreira et al. (2006a), measuring, at least for the outer portion of the jet, a large launching disc radius for a sample of T-Tauri stars (but see caveats on the estimate of the disc radii discussed in e.g. Cai 2009). 
Anderson et al. (2003) and Ferreira et al. (2006a) also suggest the possible existence of a coaxial, faster flow inside the outer jet which would account for the transient faster blobs observed in YSOs. This second flow might be powered by one of the other two mechanisms, stellar-wind or X-wind (see Ferreira et al. 2006a for caveats on the models and a detailed discussion).

Note that also in NS XRBs, a faster, relativistic jet superimposed on a slower radio ejection has been claimed in two NS systems (Sco~X-1 and Cir~X-1; Fomalont et al. 2001; Fender et al. 2004) and suggested also in SS~433 (Migliari et al. 2005; where, although recent works point towards a BH accretor, the nature of the compact object is not yet firmly established: e.g. Kubota et al. 2010). In these cases, the jet may tap power from the angular momentum of the compact object itself, and the spin and stellar magnetic field may, in fact, be crucial elements.

An active role of the compact object in the formation of jets would imply a  coupled role of spin and magnetic field, as the existence and characteristics of key parameters such as e.g. the co-rotational radius in the X-wind model or the propeller-driven regime, would depend ultimately on both the spin and the magnetic field (for a given NS and mass accretion rate). Little is known, however, about the NS magnetic field of XRBs. Estimates are possible only for high-magnetic field NSs showing cyclotron lines or for AMXPs, through studies of the disc-magnetosphere interactions (e.g. Psaltis \& Chakrabarti 1999). The estimates of the magnetic field of AMXPs are usually around $\sim10^{9}$~G. In the literature, it is often assumed that non-pulsating XRBs would have lower magnetic fields than those of AMXPs, of $\sim10^{8}$~G, but the assumption is not backed by any direct observation. A thorough discussion of the role of the magnetic fields in jet production is therefore not possible based only on the sample presented in this work. Note, however, that in Fig.~1, AMXPs seem to cluster at a higher radio luminosity than normal atolls in the hard state, for a given X-ray luminosity. 

We speculate the existence of a second process which produces coaxial, faster jets during high-accretion rate regimes, and where the characteristics of the compact object play a significant role (Option {\it iv}). This second process, which should be dependent on the accretion regime of the system, might be the cause of 1) the radio emission observed in atoll sources in their soft state (with an X-ray luminosity well above $\sim10\%$ Eddington: 4U1820-34 and Ser X-1); 2) the systematic stronger radio emission from AMXPs during outburst (when the disc is closer to the compact object and interacts with the stronger NS magnetosphere) with respect to atolls in the hard state at a comparable X-ray luminosity and  3) the hidden relativistic jet inferred in highly accreting NSs. 

Many possible jet/wind mechanisms which strongly depend on the magnetic field and spin of the accreting star have been proposed. A possibility is the centrifugal ejection in propeller-driven outflows (e.g. Romanova et al. 2005), as suggested by David Meier (private communication) if the source is in a propeller regime. Another scenario might be one where the centrifugally driven magnetic field flows from a narrow region of the innermost disc. Two possible mechanisms with these characteristics are the Reconnection X-winds (Ferreira et al. 2000) if the disc has a high vertical magnetic field and the star rotates faster than the disc, or the X-winds (Shu et al. 1994), working (but see caveats in Ferreira et al. 2006a) if the Keplerian angular velocity of the disc at the disc-magnetosphere interaction region matches the spin of the rotating star (see also Campbell 2010). Another possibility is the multi-flow configuration proposed by Ferreira et al. (2006b) for BH XRBs, where a slower magnetohydrodynamic  `extended disc jet' coexists with a faster, relativistic  $e^{+}-e^{-}$ pair beam within the jet axis  (Henri \& Pelletier 1991). This ejection mechanism would be directly dependent on the accretion regime and, in the specific case of BH XRBs, it would be triggered around the hard-to-soft state transition, during the luminous intermediate state at high mass accretion rates.

\section{Conclusions}

This work is a first step in trying to investigate, on a quantitative footing, the possible influence of the spin on jet power in NS X-ray binaries. The study is motivated by the significant implications that a constraining result, either way, would have in models for jet formation and extraction of energy and angular momentum from a compact object. 
We collected all the NS XRBs with a measurement of the spin frequency and a detection of the radio or IR jet. As an estimate of the jet power ranking between sources, we use the normalisation of the power law that fits the X-ray/radio or X-ray/IR luminosity correlations, i.e. given an X-ray luminosity (proxy of the accreting power), the higher the normalisation the higher the jet power. 

While our use of the normalisation of the luminosity correlations as a measure of the ranking jet power is subject to large uncertainties, no better proxy for the jet power is available. However, we urge caution in over-interpreting the spin-jet power correlations, particularly given the strong dependence of our result on the (highly uncertain) assumed power law index of the luminosity correlations (which span less than 3 orders of magnitude in X-ray luminosity). 

Under these assumptions, the main results can be summarized as follows. 

1) We find a hint for a relation between spin frequency and jet power, if we fit the X-ray/radio luminosity correlation using a power law with slope 1.4, as observed in 4U 1728-34 and as predicted for a radiatively efficient disc and a total jet power proportional to the mass accretion rate.  If we use a slope of 0.6, as observed in Aql X-1, no significant relation is found.  

2) There is an indication for a similar positive relation between spin frequency and jet power within the class of AMXPs alone, if we fit the X-ray/infrared luminosities using a power law with slope 1.4.

Due to the present uncertainties, further observations are needed to constrain the X-ray/radio luminosity correlations  and confirm the reliability of the jet power ranking estimation method in NSs, and to test the results of this work. In our opinion, further studies should especially focus on AMXPs for which a precise measurement of the spin is available and the magnetic field strength has been estimated, therefore enabling us to disentangle the coupled role of spin and magnetic field. Further, deeper studies of outburst cycles in atolls and AMXPs with new X-rays, radio and IR monitoring campaigns will be able to test the correlations with  much better statistics, extending the luminosity range of up to an order of magnitude. 
NS XRBs, which host a relativistic, magnetized, accreting object, are crucial for advancing our understanding of jet formation in general, building a bridge between the two most studied jet-producing systems: BHs, i.e. non-magnetized, relativistic objects; and YSOs, i.e. non-relativistic, magnetized stars.

\section*{Acknowledgments}

SM acknowledges financial support from MEC and European Social Funds through a Ram\'on y Cajal and the Spanish MICINN through grant AYA2010-21782-C03-01. DMR acknowledges support from a Netherlands Organisation for Scientific Research (NWO) Veni Fellowship.

{}


\begin{thebibliography}{}


\bibitem[\protect\citeauthoryear{}{}]{} Altamirano D.,  Casella P., Patruno A., Wijnands R.,  \& van der Klis M.,  2008, ApJL, 674, L45 

\bibitem[\protect\citeauthoryear{}{}]{} Anderson J.~M., Li  Z.-Y., Krasnopolsky R., \& Blandford R.D., 2003, ApJL, 590, L107

\bibitem[\protect\citeauthoryear{}{}]{} Blandford, R.~D., \& Payne, D.~G., 1982, MNRAS, 199, 883 

\bibitem[\protect\citeauthoryear{}{}]{} Blandford, R.~D., \& Znajek, R.~L., 1977, MNRAS, 179, 433 

\bibitem[\protect\citeauthoryear{}{}]{} Bradshaw, C.~F.,  Geldzahler, B.~J., \& Fomalont, E.~B., 2003, ApJ, 592, 486 

\bibitem[\protect\citeauthoryear{}{}]{} Cai, M.J., 2009, A\&SS Proceedings, in Protostellar Jets in Context, Part 3, 143

\bibitem[\protect\citeauthoryear{}{}]{} Casella, P.,  Altamirano, D., Patruno, A., Wijnands, R.,  \& van der Klis, M.,  2008, ApJ, 674, L41 

\bibitem[\protect\citeauthoryear{}{}]{} Chakrabarty, D.,  Morgan, E.~H., Muno, M.~P., Galloway, D.~K., Wijnands, R., van der Klis,  M., \& Markwardt, C.~B., 2003, Nature, 424, 42 

\bibitem[\protect\citeauthoryear{}{}]{} Corbel, S., Fender R.P., 2002, ApJ, 573, L35

\bibitem[\protect\citeauthoryear{}{}]{} Corbel S. et al., 2003, A\&A, 400, 1007

\bibitem[\protect\citeauthoryear{}{}]{} Coriat, M., et al., 2011, MNRAS accepted, arXiv:1101.5159

\bibitem[\protect\citeauthoryear{}{}]{} Coriat, M., Corbel, S., Buxton, M. M., Bailyn C. D., Tomsick J. A., K\"ording, E., Kalemci, E., 2009, MNRAS, 400, 123

\bibitem[\protect\citeauthoryear{}{}]{} Falcke, H., K\"ording, E., \& Markoff S., 2004, A\&A, 414, 895

\bibitem[\protect\citeauthoryear{}{}]{} Fender R.P., Hendry M.A., 2000, MNRAS, 317, 1

\bibitem[\protect\citeauthoryear{}{}]{} Fender R.P., Belloni T., Gallo E., 2004a, MNRAS, 355, 1105

\bibitem[\protect\citeauthoryear{}{}]{} Fender, R.P., Wu, K.,  Johnston, H., Tzioumis, T., Jonker, P., Spencer, R.,  \& van der Klis, M.,  2004, Nature, 427, 222 

\bibitem[\protect\citeauthoryear{}{}]{} Fender, R.P., De Bruyn, G., Pooley, G., Stappers, B., 2004c, Astronomer's Telegrams, 361,1  

\bibitem[\protect\citeauthoryear{}{}]{} Fender, R.~P., Dahlem,  M., Homan, J., Corbel, S., Sault, R.,  \& Belloni, T.~M., 2007, MNRAS, 380, L25 

\bibitem[\protect\citeauthoryear{}{}]{} Fender, R. P., Homan, J., Belloni, T. M., 2009, MNRAS, 396, 1370

\bibitem[\protect\citeauthoryear{}{}]{} Fender, R.~P., Gallo,  E., \& Russell, D., 2010, MNRAS, 406, 1425 

\bibitem[\protect\citeauthoryear{}{}]{} Ferreira J., Pelletier G., Appl S., 2000, MNRAS, 312, 387

\bibitem[\protect\citeauthoryear{}{}]{} Ferreira J., Dougados C., \& Cabrit S., 2006a, A\&A, 453, 785

\bibitem[\protect\citeauthoryear{}{}]{} Ferreira J., Petrucci P-O, Henri G., Saug\'e L., Pelletier G., 2006b, A\&A, 447, 813

\bibitem[\protect\citeauthoryear{}{}]{} Fomalont, E.~B.,  Geldzahler, B.~J., \& Bradshaw, C.~F.,  2001, ApJ, 558, 283 

\bibitem[\protect\citeauthoryear{}{}]{} Gallo E., Fender R.P., Pooley G.G., 2003, MNRAS, 344, 60

\bibitem[\protect\citeauthoryear{}{}]{} Gaensler B.M., Stappers B.W., Getts T.J., 1999, ApJ, 522, L117

\bibitem[\protect\citeauthoryear{}{}]{} Giles A. B., Greenhill J. G., Hill K. M., Sanders E., 2005, MNRAS, 361, 1180

\bibitem[\protect\citeauthoryear{}{}]{} Greenhill J. G., Giles A. B., Coutures C., 2006, MNRAS, 370, 1303

\bibitem[\protect\citeauthoryear{}{}]{} Hasinger, G. \& van der Klis, M., 1989, A\&A, 225, 79

\bibitem[\protect\citeauthoryear{}{}]{} Henri G. \& Pelletier G., 1991, ApJ, 383, L7 

\bibitem[\protect\citeauthoryear{}{}]{} Hjellming, R.~M., Han, X.~H., Cordova, F.~A., \& Hasinger, G., 1990a, A\&A, 235, 147 

\bibitem[\protect\citeauthoryear{}{}]{} Hjellming, R.~M., et  al.,  1990b, ApJ, 365, 681 

\bibitem[\protect\citeauthoryear{}{}]{} Homan, J., et al.,  2010,  ApJ, 719, 201 

\bibitem[\protect\citeauthoryear{}{}]{} Jonker, P. G.; Nelemans, G., 2004, MNRAS, 354, 355

\bibitem[\protect\citeauthoryear{}{}]{}  Jonker, Peter G.; van der Klis, Michiel; Wijnands, Rudy; Homan, Jeroen; van Paradijs, Jan; M\'endez, Mariano; Ford, Eric C.; Kuulkers, Erik; Lamb, Frederick K., 2000, ApJ, 537, 374

\bibitem[\protect\citeauthoryear{}{}]{} Kato S., 2004, PASJ, 56, L25

\bibitem[\protect\citeauthoryear{}{}]{} K\"onigl, A., \& Pudritz, R.~E., 2000, Protostars and Planets IV, 759 

\bibitem[\protect\citeauthoryear{}{}]{} K\"ording, E.~G.,  Fender, R.~P., \& Migliari, S.,  2006, MNRAS, 369, 1451 

\bibitem[\protect\citeauthoryear{}{}]{} K\"ording, E., Rupen, M., Knigge, C., Fender, R., Dhawan, V., Templeton, M., Muxlow, T., 2008, Science, 320, 1318

\bibitem[\protect\citeauthoryear{}{}]{} K\"ording, E. G.; Migliari, S.; Fender, R.; Belloni, T.; Knigge, C.; McHardy, I., 2007, MNRAS, 380, 301

\bibitem[\protect\citeauthoryear{}{}]{} Krauss M. I., et al., 2005, ApJ, 627, 910

\bibitem[\protect\citeauthoryear{}{}]{} Kubota, K.; Ueda, Y.; Fabrika, S.; Medvedev, A.; Barsukova, E. A.; Sholukhova, O.; Goranskij, V. P., ApJ, 2010, 709, 1374

\bibitem[\protect\citeauthoryear{}{}]{} Lin, D., Remillard, R. A., Homan, J., 2009, ApJ, 696, L1257

\bibitem[\protect\citeauthoryear{}{}]{} Linares, M., van der Klis, M., Wijnands, R., 2007, ApJ, 660, 595  

\bibitem[\protect\citeauthoryear{}{}]{} Lewis F., et al., 2010, A\&A, 517, A72

\bibitem[\protect\citeauthoryear{}{}]{} Maitra D., Bailyn C.D., ApJ, 2004, 608, 444

\bibitem[\protect\citeauthoryear{}{}]{} Merloni, A., Heinz, S.,  \& di Matteo, T., 2003, MNRAS, 345, 1057 

\bibitem[\protect\citeauthoryear{}{}]{} Migliari, S., \& Fender, R.~P.,  2006, MNRAS, 366, 79 

\bibitem[\protect\citeauthoryear{}{}]{} Migliari, S., Fender,  R.~P., Rupen, M., Jonker, P.~G., Klein-Wolt, M., Hjellming, R.~M.,  \& van der Klis, M., 2003, MNRAS, 342, L67 

\bibitem[\protect\citeauthoryear{}{}]{} Migliari, S., Fender,  R.~P., Rupen, M., Wachter, S., Jonker, P.~G., Homan, J.,  \& van der Klis, M., 2004, MNRAS, 351, 186 

\bibitem[\protect\citeauthoryear{}{}]{} Migliari, S., Fender,  R.~P., Blundell, K.~M., M{\'e}ndez, M.,  \& van der Klis, M., 2005, MNRAS, 358, 860 

\bibitem[\protect\citeauthoryear{}{}]{} Migliari, S., et al.,  2007, ApJ, 671, 706 

\bibitem[\protect\citeauthoryear{}{}]{} Migliari, S., et al.,  2010, ApJ, 710, 117 

\bibitem[\protect\citeauthoryear{}{}]{}  Miller-Jones, J.~C.~A., et al., 2010, ApJL, 716, L109 

\bibitem[\protect\citeauthoryear{}{}]{} Miller-Jones, J.~C.~A., et al., 2011, Jets at all Scales, Proceedings of the International Astronomical Union, IAU Symposium, 275, 224

\bibitem[\protect\citeauthoryear{}{}]{} Moore, C.~B., Rutledge,  R.~E., Fox, D.~W., Guerriero, R.~A., Lewin, W.~H.~G., Fender, R.,  \& van Paradijs, J.,  2000, ApJ, 532, 1181 

\bibitem[\protect\citeauthoryear{}{}]{} Penninx, W., Lewin,  W.~H.~G., Zijlstra, A.~A., Mitsuda, K.,  \& van Paradijs, J., 1988, Nature, 336, 146 

\bibitem[\protect\citeauthoryear{}{}]{} Pooley G.G., 2004, ATEL, 355

\bibitem[\protect\citeauthoryear{}{}]{} Psaltis, D., \& Chakrabarty, D., 1999, ApJ, 521, 332 

\bibitem[\protect\citeauthoryear{}{}]{} Reig, P., van Straaten,  S., \& van der Klis, M.,  2004, ApJ, 602, 918 

\bibitem[\protect\citeauthoryear{}{}]{} Romanova, M. M.; Ustyugova, G. V.; Koldoba, A. V.; Lovelace, R. V. E., 2005, ApJ, 635, L165

\bibitem[\protect\citeauthoryear{}{}]{} Remillard R. A., McClintock J. E., 2006, ARA\&A, 44, 49

\bibitem[\protect\citeauthoryear{}{}]{} Rupen M.P., Dahwan V., Mioduszewski A.J., 2002, IAU Circ., 7893

\bibitem[\protect\citeauthoryear{}{}]{} Rupen M.P., Dahwan V., Mioduszewski A.J., 2002a,  The Astronomer's Telegram, 524

\bibitem[\protect\citeauthoryear{}{}]{} Rupen, M.P., Dhawan, V., Mioduszewski, A. J., 2002b, IAU Circ., 7893

\bibitem[\protect\citeauthoryear{}{}]{} Russell, D. M.; Fender, R. P.; Hynes, R. I.; Brocksopp, C.; Homan, J.; Jonker, P. G.; Buxton, M. M., 2006, MNRAS, 371, 1334

\bibitem[\protect\citeauthoryear{}{}]{} Russell D. M., Fender R. P., Jonker P. G., 2007, MNRAS, 379, 1108

\bibitem[\protect\citeauthoryear{}{}]{} Russell, D. M.; Fender, R. P.; Jonker, P.; Maitra, D., 2008, AIPC,  1068, 221

\bibitem[\protect\citeauthoryear{}{}]{} Russell D.M.,  Fender R.P., 2010, "Black Holes and Galaxy Formation", Nova Science Publishers, Inc., at press:  arXiv:1001.1244

\bibitem[\protect\citeauthoryear{}{}]{} Rutledge, R., Moore,  C., Fox, D., \& Lewin, W., 1998, The Astronomer's Telegram, 8, 1 

\bibitem[\protect\citeauthoryear{}{}]{} Sell, P.~H., et al., 2010,  ApJ, 719, L194 

\bibitem[\protect\citeauthoryear{}{}]{} Shu, F., Najita, J.,  Ostriker, E., Wilkin, F., Ruden, S., \& Lizano, S., 1994, ApJ, 429, 781 

\bibitem[\protect\citeauthoryear{}{}]{} Soleri, P., Tudose, V.,  Fender, R., van der Klis, M., \& Jonker, P.~G.,  2009, MNRAS, 399, 453 

\bibitem[\protect\citeauthoryear{}{}]{} Soleri P., Fender R.P., 2011, MNRAS accepted, arXiv:1101.1214

\bibitem[\protect\citeauthoryear{}{}]{} Strohmayer, T.~E.,  Markwardt, C.~B., Swank, J.~H., \& in't Zand, J., 2003, ApJL, 596, L67

\bibitem[\protect\citeauthoryear{}{}]{} Tan, J., Lewin, W.~H.~G.,  Hjellming, R.~M., Penninx, W., van Paradijs, J., van der Klis, M.,  \& Mitsuda, K., 1992, ApJ, 385, 314 

\bibitem[\protect\citeauthoryear{}{}]{} Torres M. A. P., et al., 2008, ApJ, 672, 1079

\bibitem[\protect\citeauthoryear{}{}]{} Tudose V., Migliari S., Miller-Jones J.C.A., Nakajima M., Yamaoka K., Kuulkers E., 2010, The Astronomer's Telegram, 2798, 1

\bibitem[\protect\citeauthoryear{}{}]{} Tudose, V., Fender, R.~P., Linares, M., Maitra, D.,  \& van der Klis, M., 2009, MNRAS, 400, 2111

\bibitem[\protect\citeauthoryear{}{}]{} Valentim, R.; Rangel, E.; Horvath, J. E., 2011, MNRAS submitted, arXiv:1101.4872

\bibitem[\protect\citeauthoryear{}{}]{} van der Klis, M., 2006, in ``Compact Stellar X-Ray Sources'', eds. W.H.G. Lewin and M. van der Klis, Cambridge University Press

\bibitem[\protect\citeauthoryear{}{}]{} van Straaten, S.,  van der Klis, M., \& Wijnands, R., 2005, ApJ, 619, 455 

\bibitem[\protect\citeauthoryear{}{}]{} Wang Z., et al., 2001, ApJ, 563, L61

\bibitem[\protect\citeauthoryear{}{}]{} Watts, A. L., Krishnan, B., Bildsten, L.,  Schutz, B. F., 2008, MNRAS, 389, 839

\end{thebibliography}
\end{document}